\documentclass[showpacs,preprintnumbers,amsmath,amssymb,superscriptaddress]{revtex4}

\usepackage{color,epsfig} 
\usepackage{graphicx}
\usepackage{rotating}
\usepackage{hyperref}

\def\be{\begin{equation}}
\def\ee{\end{equation}}
\def\baray{\begin{eqnarray}}
\def\earay{\end{eqnarray}}

\numberwithin{equation}{section}

\begin{document}
\preprint{}

\title{Constraints on Brane Inflation and Cosmic Strings}
\author{Richard A. Battye}
\affiliation{Jodrell Bank Centre for Astrophysics, University of Manchester, Manchester M13 9PL, U.K.}
\author{Bj\"orn Garbrecht}
\affiliation{Department of Physics, University of Wisconsin, Madison, WI 53706, U.S.A.}
\author{Adam Moss}
\affiliation{Department of Physics and Astronomy, University of British Columbia, Vancouver V6T 1Z1, Canada}
\author{Horace Stoica}
\affiliation{Blackett Laboratory, Imperial College, Prince Consort Road, London SW7 2BZ, U.K.}

\date{\today}
 
\begin{abstract}
By considering simple, but representative, models of brane inflation from a single brane-antibrane pair in the slow roll regime, we provide constraints on the parameters of the theory imposed by measurements of the CMB anisotropies by WMAP including a cosmic string component. We find that inclusion of the string component is critical in constraining parameters. In the most general model studied, which includes an inflaton mass term, as well as the brane-antibrane attraction, values $n_{\rm s}<1.02$ are compatible with the data at $95\%$ confidence level. We are also able to constrain the 
volume of the warped throat region (modulo factors dependent on the warp factor) and the value of the inflaton field to be $<0.66M_{\rm P}$ at horizon exit. We also investigate models with a mass term. These observational considerations suggest that such models have $r< 2\times 10^{-5}$, which can only be circumvented in the fast roll regime, or by increasing the number of antibranes. Such a value of $r$ would not be detectable in any CMB polarization experiment likely in the near future, but the B-mode signal from the cosmic strings could be detectable. We present forecasts of what a similar analysis using PLANCK data would yield and find that it should be possible to rule out $G\mu > 6.5\times 10^{-8}$ using just the TT, TE and EE power spectra.
\end{abstract}

\pacs{11.25.-w, 98.80.Cq, 98.80.Es}

\maketitle

\section{Introduction}

The inflationary paradigm is strongly supported by observations of the cosmic microwave background (CMB) made by the COBE and WMAP satellites~\cite{Smoot:1992td,Spergel:2006hy}. However, it is still a paradigm in search of a specific model based on fundamental physics. Brane inflation~\cite{Dvali:1998pa} has emerged as one of the most popular ways of embedding inflation within string theory. It uses a natural candidate for the inflaton: the field which describes the brane-antibrane separation. This field has a non-trivial potential due to the attractive brane-antibrane interaction which is flattened by the effect of the compactification of the extra dimensions \cite{Burgess:2001fx}, and the geometry of the branes \cite{GarciaBellido:2001ky}.

Cosmic strings are also a natural occurrence within this model and are result of inhomogeneities in the tachyon field of the brane-antibrane pair \cite{C_Strings1,C_Strings2}. These are represented by a complex field with a non-trivial potential which supports the formation of codimension 2 defects which have been shown to exhibit the correct properties of lower dimension branes \cite{Tachyon_Strings}. Since reheating of the universe in these model proceeds via the annihilation of the brane-antibrane pair, the formation of cosmic strings is expected at the end of the inflation and the strings will be naturally located at the bottom of the throat as their tension is also warp-factor dependent. 

The most complete model of brane inflation incorporating moduli stabilization has been proposed in ref.~\cite{Kachru:2003sx}. In this model inflation happens naturally as a mobile brane falls down the warped throat, being attracted by an antibrane stuck at the bottom of the throat. Antibranes have a warp-factor dependent potential, and therefore minimize their energy by moving to the bottom of the throat,  the region of strongest warping. The mobile brane is not affected by the warping. One can understand this by considering the warped geometry as being generated by a large stack of branes. An antibrane is attracted to the stack and will move towards it, that is towards largest warping. The brane feels no force from the stack of branes (it is BPS with respect to them), only 
from the antibrane located at the bottom of the throat. This situation can be described by a simple scalar field, since in this model all moduli are stabilized and only the brane-antibrane separation is evolving.  

Constraints on the cosmic string tension, $G\mu$, come from a variety of observations. Of most interest here are those which are most robust. In particular, we will concentrate on the constraints which result from their inclusion as a sub-dominant component in the angular power spectrum of anisotropies of the cosmic microwave background (CMB)~\cite{CHMb,WB,Wyman:2005tu,Seljak:2006bg,Bevis:2006mj,Battye:2006pk,Bevis:2007gh}. As pointed out recently~\cite{Battye:2006pk,Bevis:2007gh} in the context of the third year WMAP data, larger values of the spectral index of density fluctuations, $n_{\rm s}$, are compatible with observations if a sub-dominant string component with around $5-10\%$ of the large-scale amplitude is included. This is a generic feature of any inflation model which produce strings. In ref.~\cite{Battye:2006pk} accurate constraints on the coupling constant, $\kappa$ and mass scale, $M$, relevant to the simplest models of supersymmetric F- and D-term hybrid inflation were computed, taking into account the fact that the observed power spectrum is described by 3 parameters, $G\mu$, $n_{\rm s}$ and the power spectrum amplitude, $P_{\cal R}$, each of which can be derived from $\kappa$ and $M$. In this model dependent approach more powerful constraints are possible. We will adapt the same approach, where relevant to the case of brane inflation in this paper. 

\section{Details of brane inflation}

\subsection{Simplest case}

The simplest possible inflaton potential which can lead to brane inflation is 
\begin{equation}
\label{simppot}
V=V_0\left(1-\frac{\gamma}{\phi^4}\right)\,,
\end{equation}
which corresponds to the attractive interaction between a $D3$ brane and an antibrane in  extra-dimensions, when the compactification manifold is a 6-torus or when the brane-antibrane separation is small enough that the effects of the moduli stabilization can be neglected. 
The canonically normalized inflaton $\phi$ is proportional to the distance between these branes.
In this situation, there are only two parameters, namely $V_0$ and $\gamma$.

The parameters $V_0$ and $\gamma$ can be linked directly to the fundamental 
picture proposed in ref.~\cite{Kachru:2003sx}. $V_0$ can be expressed as
\begin{equation}
V_0=\frac{M M_{\rm S}^4 h_A^4}{2(2\pi)^3 g_{\rm s}}
=2 M T_3 h_A^4\,.
\end{equation}
Here $h_A$ is the warp-factor at the position of the antibrane, $M_{\rm S}$ is the string-scale,  $M$ is the number of colliding brane-antibrane pairs (which we will set to 1 unless otherwise stated in this paper, corresponding to a single pair),  $g_s$ the fundamental string coupling, which we will set to $2\alpha_{\rm GUT}\approx 2/25$ the value suggested by gauge-coupling unification, and
$T_3$ is the tension of the 3-brane.
We note that in terms of these fundamental parameters, the brane separation
is given by $\phi/\sqrt{3 M T_3}$. Within the expression
\begin{equation}
\gamma=36 M_{\rm S}^4 M^3 h_A^4 \frac{\pi a}{(2 \pi)^6g_{\rm S}}\,,
\label{gamma}
\end{equation}
a new parameter $a$ is introduced. It is related to the volume of
the warped throat region $X_5$ by
\begin{equation}
a=\frac{\pi^{3}}{\text{Vol} X_{5}}\left\{
\begin{array}{ccc}
=1 & \text{for} & S_{5} \\
=\frac{27}{16} & \text{for} & T^{1,1} \\
\end{array}
\right.\,,
\label{vol}
\end{equation}
where the metric of the throat is
given by ${\rm AdS}_5\times X_5$. The minimal value of $a$ is obtained for a 5-sphere, but large values for $a$ can be obtained by considering internal manifolds with a large-rank orbifold group. The most popular model is the $T^{1,1}$ manifold considered in ref.~\cite{Klebanov:2000hb}, which has volume $\text{Vol} ~T^{1,1} = 16\pi^3/27$.

The quantities relevant for observation can be found by simple analytical calculations~\cite{Firouzjahi:2005dh} using slow-roll approximation, and we summarize the results here. The value of the inflaton at horizon exit of the fiducial
scale $k=0.05{\rm Mpc}^{-1}$ in terms of the number of
e-foldings $N_{\rm e}$, which this scale experiences during inflation,
is given by
$\phi_{\rm e}\approx (24 \gamma N_{\rm e} M_{\rm P}^2)^{\frac16}
-(20\gamma M_{\rm P}^2)^\frac16\,$, where $M_{\rm P}= (8\pi G)^{-1/2}$ is the reduced Planck mass. The precise number of e-foldings depends weakly on the scale of inflation and also on the thermal history of the Universe. Here, we set $ N_{\rm e}=53$ which corresponds to a reheat temperature around $10^{16}{\rm GeV}$ within the standard cosmological scenario. This could be included as a parameter which is marginalized over, but the results will only be slightly modified.

At the fiducial scale, one then finds the amplitude of scalar perturbations
\begin{equation}
\sqrt{P_{\cal R}}\approx 
\frac{ \sqrt{V_0} (24 N_{\rm e} +20)^{\frac56}}{8\pi\sqrt{3}M_{\rm P}^{4/3}\gamma^{1/6}}\approx 9.0\left({V_0\over M_{\rm P}^4}\right)^{1/2}\left({\gamma\over M_{\rm P}^4}\right)^{-1/6}\,,\label{amplitude}
\end{equation}
the scalar spectral index
\begin{eqnarray}
n_{\rm s}\approx 1-\frac{40}{24 N_{\rm e}+20}-
48\frac{\gamma^{1/3}}{M_{\rm P}^{4/3}
(24 N_{\rm e} +20)^{\frac{10}{6}}}\approx 0.97-3.1\times 10^{-4}\left({\gamma\over M_{\rm P}^4}\right)^{1/3} \label{specindex}
\,,
\end{eqnarray}
and the tensor-to-scalar ratio
\begin{equation}
r\approx 128\frac{\gamma^{1/3}}{M_{\rm P}^{4/3}(24 N_{\rm e} +20)^{\frac{10}{6}}}\approx 8.4\times 10^{-4}\left({\gamma\over M_{\rm P}^4}\right)^{1/3}\,.\label{scalartensor}
\end{equation}

The main focus of this paper are the signatures of the
network of $D$-strings in the anisotropy spectrum of the CMB. As mentioned already, the strings form at the end of inflation, when the branes collide, and their tension is given by
\begin{equation}
\label{stringtension}
G\mu=\frac{\sqrt{V_0}}{8\sqrt{M\pi g_s}M_{\rm P}^2}\approx 0.25\left({V_0\over M_{\rm P}^4}\right)^{1/2} \,.
\end{equation}
using $M=1$ and $g_s=2/25$. Therefore, the string tension  does not depend on more parameters than already introduced in the potential~(\ref{simppot}). We will use the string spectrum used in our earlier work~\cite{Battye:2006pk} which derives from refs.~\cite{VP,ABR}, in addition to standard adiabatic spectra computed using  codes such as {\tt CMBFAST}~\cite{Seljak} and {\tt CAMB}~\cite{LC}.

\subsection{Inclusion of a confining mass term}

The simple potential (\ref{simppot}) can be augmented by the inclusion of a variety of positive powers of $\phi$ in order include extra effects from the background geometry. In particular it has been argued~\cite{Kachru:2003sx} that moduli stabilization~\cite{Kachru:2003aw} will typically lead to a warped geometry and therefore the antibrane, having a warp factor dependent tension, will be  located at the bottom of the warped throat. While the mobile brane is not affected by the fluxes generating the
throat geometry it is affected by the volume stabilization mechanism resulting in a non-zero mass term for the inflaton. Therefore we will consider more general potentials of the type
\begin{equation}
\label{masspot}
V=V_0\left({\beta\over 6M_{\rm P}^2} \phi^2
+1-\frac{\gamma}{\phi^4}\right)\,.
\end{equation}
Such scenarios require some mechanism, or fine-tuning, to
remove other possible terms in the potential; notably the linear term. It will also require $\beta$ tuned to be $<1$ to achieve slow roll.

The simplest string theory realizations of this potential predict values of $\beta\sim{\cal O}(1)$ leading to a new $\eta$-problem~\cite{Kachru:2003sx}, with the slow-roll parameter $\eta\propto V^{\prime\prime}/V $ being too large to allow slow-roll. In ref.~\cite{McAllister:2005mq}, it was pointed out that both F-term and D-term inflation models suffer from this $\eta$ problem when the stabilization of the volume modulus is achieved via non-perturbative effects generating a
non-trivial superpotential. A small mass for the inflaton can still be
obtained, but this will require a certain level of tuning of the parameters. 

In F-term models the stabilizing potential for the volume modulus is
the F-term potential itself. Hence, the inflaton always appears
in the  K\"ahler potential~\cite{DeWolfe:2002nn} and as a result, the same effects  that stabilize the volume modulus generate a mass for the inflaton. 
Generically, the masses are of the order of the string mass, and hence $\eta\sim{\cal O}(1)$. Inflation can still be obtained in these
models, but a fine-tuning is required~\cite{Burgess:2004kv}. 

For D-term inflation models the K\"ahler potential does not appear
in the D-term, so naively one expects that no mass term 
will be generated for the inflaton. However, in specific
$D3-D7$ inflationary models, a
vanishing inflaton mass is expected as a result of a shift symmetry. 
Such models typically contain $D7$ branes in which gaugino condensation 
takes place and therefore threshold (one loop) corrections  
~\cite{Berg:2004ek,Berg:2005ja,Berg:2004sj} 
to the gauge kinetic function of the 
$SU\left(N\right)$ Yang-Mills living inside the stack of $D7$ branes 
re-introduce an inflaton dependence on the potential.
Again, fine-tuning is required for inflation to work.

The inclusion of $\beta$ modifies the expressions for the observable quantities $P_{\cal R}$, $n_{\rm s}$ and $r$~\cite{Firouzjahi:2005dh}.
For comparison with results presented below, we
use the fact that $\beta\ll 1$ for viable slow-roll
inflation, and find for the scalar spectral index the
expression
\begin{equation}
\label{specindex:betanonzero}
n_{\rm s}\approx1+\frac 23 \beta - \frac {10}3 \beta
\frac 1{{\rm e}^{2 \beta N_{\rm e}}-1}\,.
\end{equation}
This reproduces the full result within slow-roll approximation, which we use
for the MCMC analysis, to a good
accuracy. Note also that this expression
reduces for $\beta \to 0$ to the formula~(\ref{specindex}),
when neglecting
the small contribution $\propto \gamma^{1/3}$.

We note that some models of brane inflation presented in the literature do not appear at first sight to have potentials of the form (\ref{masspot}) which we have claimed is the simplest form possible (for example, refs.\cite{Baumann:2007np,Baumann:2007ah}). Such models, which feature moduli stabilization, 
generically require fine-tuning in order to obtain a flat enough
potential such that inflation lasts more than 55 e-folds.
This fine-tuning is realised by balancing against each other two
potential terms with opposite curvature (second derivative)
and this leads to a potential featuring an inflexion point; it is
the region around this point where the $\eta$ parameter is small enough
that slow-roll, and therefore inflation, takes place. In ref.~\cite{Baumann:2007np} this inflexion point results from balancing the effects
of the loop corrections against those of the K\"ahler moduli stabilization.
In (\ref{masspot}) the effect of the K\"ahler moduli stabilization
(which we collectively describe via the parameter $\beta$) is balanced
against the Coulombic brane-anti-brane interaction which have
opposite curvature and the potential features an inflexion point like the
other models do. However, we prefer not to expand the potential around the
inflexion point as this will make it more difficult to relate the results
of the analysis with the parameters of the underlying model. Hence, although the forms of the potential used sometimes appear very different to (\ref{masspot}) the physical situation under consideration is very much the same.

\section{Results of MCMC analysis}

\subsection{Current constraints}

The MCMC analysis used the November 2006 version of {\tt COSMOMC}~\cite{LB} in order to create chains to estimate confidence limits on the cosmological parameters. The basic set of five parameters $\{ \Omega_{\rm b}h^2, \Omega_{\rm c}h^2, \tau_{\rm R}, \theta_{\rm A}, \log(10^{10}P_{\cal R}) \}$, where $\theta_{\rm A}$ is defined by the ratio of the sound horizon to the angular diameter distance at the redshift of recombination, were used in each case. For the simplest inflationary model (\ref{simppot}) we use $\gamma$ as an additional input parameter, deriving $\{n_{\rm s}, G\mu\}$ and related quantities such as $V_{0}$ from this and $P_{\cal R}$. There are only 6 parameters in this model. This is more sensible than using $n_{\rm s}$ as an input as only a narrow range of $n_{\rm s}$ is allowed making it difficult for {\tt COSMOMC} to locate viable models. For the more general model we use $\{ \gamma, \beta \}$ as input parameters making 7 in total. The intrinsic flat priors, listed in Table \ref{tab:flatpriors}, were chosen to be sufficiently broad to incorporate the lines of degeneracy known to exist within the space of parameters. At various points in the discussion we refer to the `standard' six-parameter fit as the set $\{ \Omega_{\rm b}h^2, \Omega_{\rm c}h^2, \tau_{\rm R}, \theta_{\rm A},\log(10^{10}P_{\cal R}),n_{s} \}$.  

\begin{table}
\begin{center}
\begin{tabular}{|c|c|} \hline
Parameter & Prior \\ \hline
$\Omega_{\rm b} h^2 $ & (0.005, 0.1) \\ 
$\Omega_{\rm c} h^2 $ & (0.01, 0.99) \\ 
$\theta_{\rm A}  $ & (0.5, 10) \\ 
$\tau_{\rm R} $ & (0.01, 0.9) \\ 
$\log (10^{10} P_{\cal R})$ & (2.7, 5.0) \\ \hline
$\log_{10}(\gamma/(10^{16}{\rm GeV})^4)$ & (-5.0, 10.0) \\
$\beta$  & (0, 0.05) \\ \hline
\end{tabular}
\end{center}
\caption{\label{tab:flatpriors} Table of flat priors. The notation $(a,b)$ for a particular parameter gives the lower and upper bounds allowed in the fit.}
\end{table}

We use data from the 3rd year observations from WMAP~\cite{Hinshaw,Page} along with the latest version of their likelihood code. The inclusion of additional CMB from other experiments does not significantly affect the constraint on $G \mu$, so we do not include this data. This also allows clearer comparison between WMAP and constraints likely to come from the next generation of CMB experiments, which we consider in the next section. Cosmic strings also provide a subdominant contribution to the galaxy power spectrum, but including this data does not improve constraints on $G\mu$ which is the main focus of this paper. 

Let us first consider the case of the simple potential (\ref{simppot}) which corresponds to the more general case (\ref{masspot}) when $\beta=0$. We have performed our analysis with and without the inclusion of the string component to the CMB power spectrum. In both cases the value of $n_{\rm s}$ is tightly constrained, not by the data, but by the restricted possibilities allowed by (\ref{specindex}). When the string component is not included, the values of $\gamma$ and $V_0$ are not strongly restricted since the power spectrum amplitude (\ref{amplitude}) is degenerate with respect to them. Any value of $\gamma$ which does not significantly effect the spectral index ($\gamma<10^{6}M_{\rm P}^4$) is allowed by the data with a corresponding weak restriction on $V_0$. Moreover, the derived value of the field $\phi_{\rm e}$ is virtually unrestricted by the data, and in particular, $\phi_{\rm e}$ may exceed the Planck scale if one does not impose a constraint. 

In the standard approach to inflation based on effective field theory, one requires that $\phi_{\rm e}<M_{\rm P}$. In this case simple arguments appear to forbid values of $r$ which are large enough to be observable within the foreseeable future~\cite{Lyth:1996im,Efstathiou:2005tq}. In brane inflation, where $\phi$ corresponds to the distance between the branes and the antibranes, large field values correspond to the infrared domain, suggesting the  possibility of bypassing the usual preconceptions against trans-Planckian field values (although see discussion of ref.~\ref{sec:strat}). Hence, if one does not include the string component, the only restrictions on $r$ can come from theoretical considerations.

If one includes the string component then things are different. The values of $\gamma$ and $V_0$ are restricted by a combination of the measured amplitude (\ref{amplitude}) and the cosmic string tension (\ref{stringtension}). We find that $\log_{10}(\gamma/(10^{16}{\rm GeV})^4)<5.3$ and $\log_{10}(V_0/(10^{16}{\rm GeV})^4)<-2.4$ at $95\%$ confidence level, corresponding to $G\mu<2.5\times 10^{-7}$. Also one finds that $\phi_{\rm e}<0.66M_{\rm P}$ at the same confidence level. We will discuss in section~\ref{sec:strat} how this leads to a bound on $r$. The constraint on $\gamma$ in conjunction (\ref{gamma}) and (\ref{vol}) implies that 
\be
{\rm Vol}X_5>1.4\times 10^{6}\left({M_{\rm S}h_{\rm A}\over 10^{16}{\rm GeV}}\right)^4\,.
\ee

\begin{figure}[t]
\begin{center}
\epsfig{file=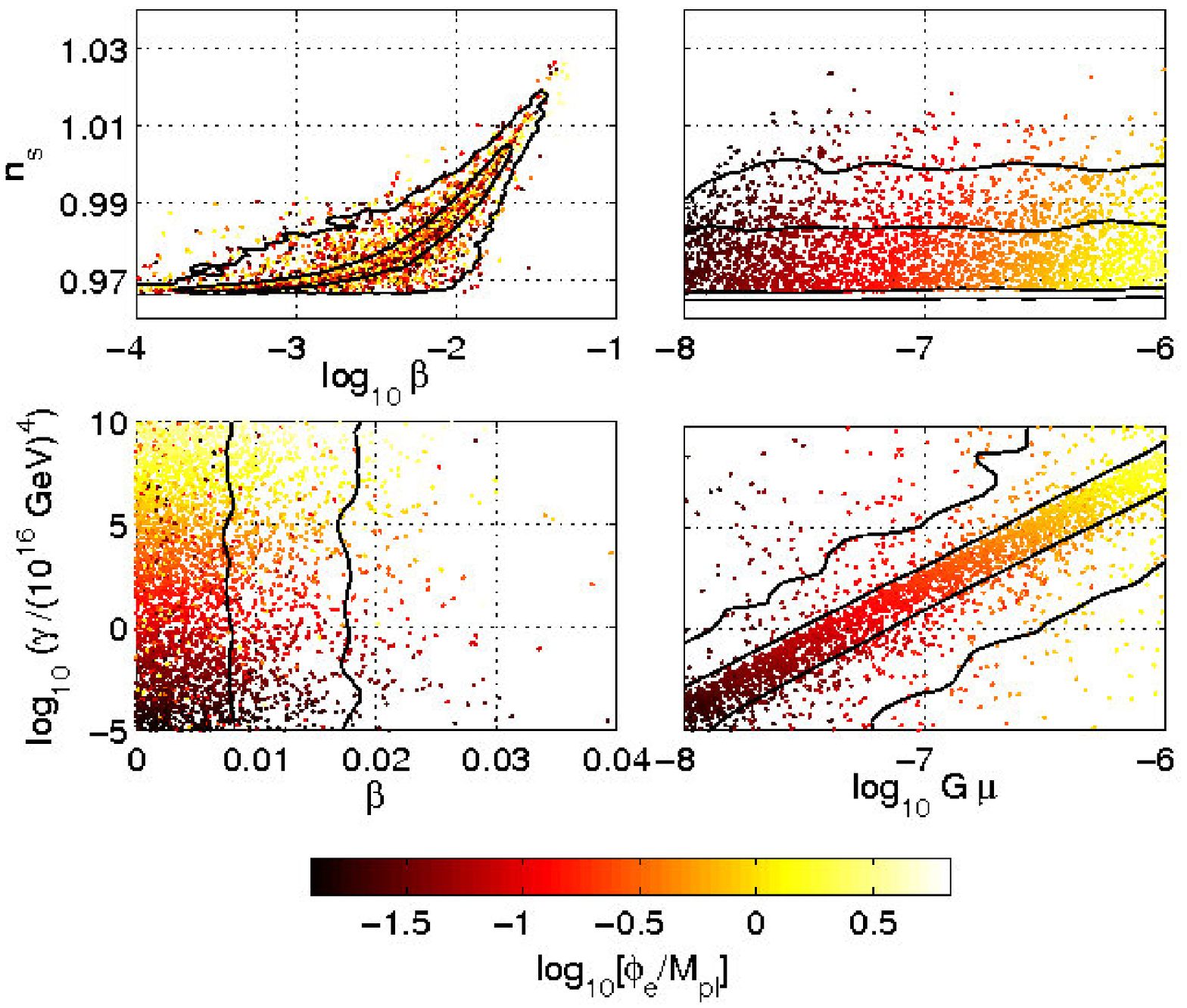,height=3.5in,width=3.5in}
\epsfig{file=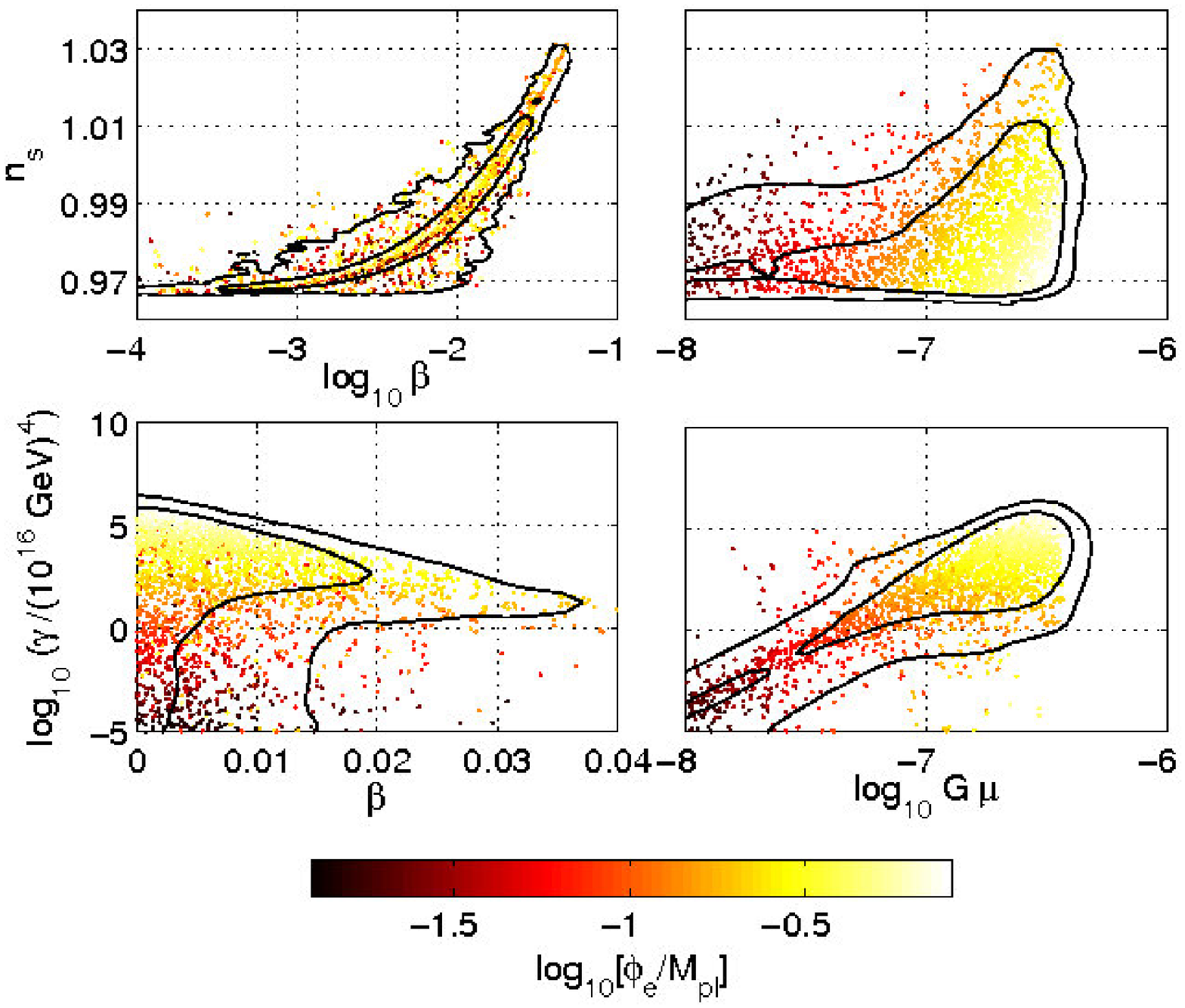,height=3.5in,width=3.5in}
\end{center}
\caption{Selected 2D likelihood surfaces for the general potential (\ref{masspot}) without (left set of four panels) and with (right set of four panels) including the cosmic string component. For each set of four: (top left) $n_{\rm s}$-$\log_{10}\beta$; (top-right) $n_s-\log_{10}G\mu$; (bottom-right) $\log_{10}(\gamma/(10^{16}{\rm GeV})^4)-\beta$; (bottom-left) $\log_{10}\gamma-\log_{10}(G\mu)$. In all cases the lines correspond to the $68\%$ and $95\%$ joint likelihood limits and the coloured dots correspond to the derived values of $\phi_{\rm e}$ for each member of the chain. It should be clear from these plots that there is an upper limit on $\beta$ and a narrow range of $n_{\rm s}$ is allowed for a specific value of $\beta$. Without including the string contribution, the value of $G\mu$ plotted is that which would be computed using (\ref{stringtension}) - there is essentially no meaningful constraint on $\gamma$ and models with $G\mu$ greater than present limits have been allowed in the analysis. With the inclusion of strings, the limit on $G\mu$ imposed by the data has no trivial effects on the allowed parameter ranges, as discussed in the text.}
\label{figure:bi_current}
\end{figure}

We have also performed the same analyses in the case of general $\beta$ and a similar qualitative picture emerges. The constraints on the parameters are summarized in Table~\ref{tab:results}, and Fig.\ref{figure:bi_current}. The important difference from the $\beta=0$ case is that a much larger range of $n_{\rm s}$ is allowed, albeit a very narrow one for each value of $\beta$ as suggested by~(\ref{specindex:betanonzero}). In this case, the constraint on $\beta$ is intimately related to that imposed by the data on $n_{\rm s}$.

\begin{table}
\begin{center}
\begin{tabular}{|c|c|c|} \hline
Parameter & Without & With \\ \hline
$\Omega_{\rm b} h^2 $ & $0.0229 \pm 0.0006$ & $0.0241 \pm 0.0014$ \\ 
$\Omega_{\rm c} h^2 $ & $0.105 \pm 0.008$ & $0.104 \pm 0.008$ \\ 
$\theta_{\rm A}  $ & $1.043 \pm 0.003$  & $1.045 \pm 0.004$ \\ 
$\tau_{\rm R} $ & $0.109 \pm 0.029$  & $0.111 \pm 0.031$ \\ 
$\log (10^{10} P_{\cal R})$ & $3.08 \pm 0.06$  & $3.05 \pm 0.07$ \\ \hline \hline
$n_{s} $ & $< 1.0$  & $< 1.02$ \\
 $h  $  & $0.75 \pm 0.03$ & $0.77 \pm 0.04$ \\ \hline \hline
$\log_{10}(\gamma/(10^{16}{\rm GeV})^4)$ & - & $< 4.9$ \\
$\beta$  & $< 0.018$ & $ < 0.025$ \\ 
$\log_{10}(V_0/(10^{16}{\rm GeV})^4) $  & -  & $< -2.3$ \\
$G \mu /10^{-7}$ & - & $< 3.1$ \\
$\phi_{\rm e}/M_{\rm P}$ & - & $< 0.56$\\ \hline \hline 
\end{tabular}
\end{center}
\caption{\label{tab:results} Constraints on the general potential (\ref{masspot}) with and without including the cosmic string component to the CMB power spectrum. The results for $n_{\rm s}, \gamma, \beta, G\mu, V_0$ and $\phi_{\rm e}/M_{\rm P}$ are marginalized $95\%$ confidence upper bounds.}
\end{table}

It is clear that the inclusion of the string component has a non-trivial effect on the allowed likelihood surfaces. In this case larger values of $n_{s}$ are allowed with a limit of $n_{\rm s}<1.02$ at $95\%$ confidence level, but with values $>1.0$ being typically requiring significant ($>10^{-7}$) values of $G\mu$. There are upper limits  $\log_{10}(\gamma/(10^{16}{\rm GeV})^4)<4.9$, $\log_{10}(V_0/(10^{16}{\rm GeV})^4)<-2.3$ and $\phi_{\rm e}<0.56 \, M_{\rm P}$.

\subsection{Future constraints}

The present constraints will be improved on by future CMB experiments, in particular the forthcoming ESA PLANCK mission~\cite{bluebook}. In order to quantify the likely improvement on the parameters discussed above, we assume that the observed CMB power spectrum is the sum of the underlying $C_{\ell}$ (which may or may not contain a string contribution) and an isotropic noise variance $N_{\ell}$. We compute the error on the estimate of $C_{\ell}$, which in turn gives confidence limits on the cosmological parameters, using the full-sky CMB likelihood function (see for example, ref.~\cite{Lewis:2006ym}). This is more accurate than Fisher matrix approaches if the posterior distribution is significantly non-Gaussian, which is certainly true for the models considered here.

We use the same noise properties as in ref.~\cite{Lewis:2006ym}, namely $N_{\ell}^{TT}=2 \times 10^{-4}$ $(\mu K)^2$ for the temperature and $N_{\ell}^{EE}=N_{\ell}^{TT}/4$ for the E-mode polarization signal, with a Gaussian beam width of 7 arcminutes. We assume an effective sky coverage $f_{\rm sky} = 0.65$,  which acts as an `fudge' factor in the full sky likelihood to approximate the partial sky case.

First, it is instructive to consider forecasted limits on $G \mu$ from PLANCK, independent of any specific inflationary model. To do this, we use a fiducial set of $C_{\ell}$'s  corresponding to the current standard 6 parameter fit  $\Lambda$CDM model~\cite{Spergel:2006hy}. For the MCMC analysis we then use $G \mu$ as an additional free parameter, and so can provide upper limits assuming the underlying model has no string contribution.

\begin{figure}[t]
\begin{center}
\epsfig{file=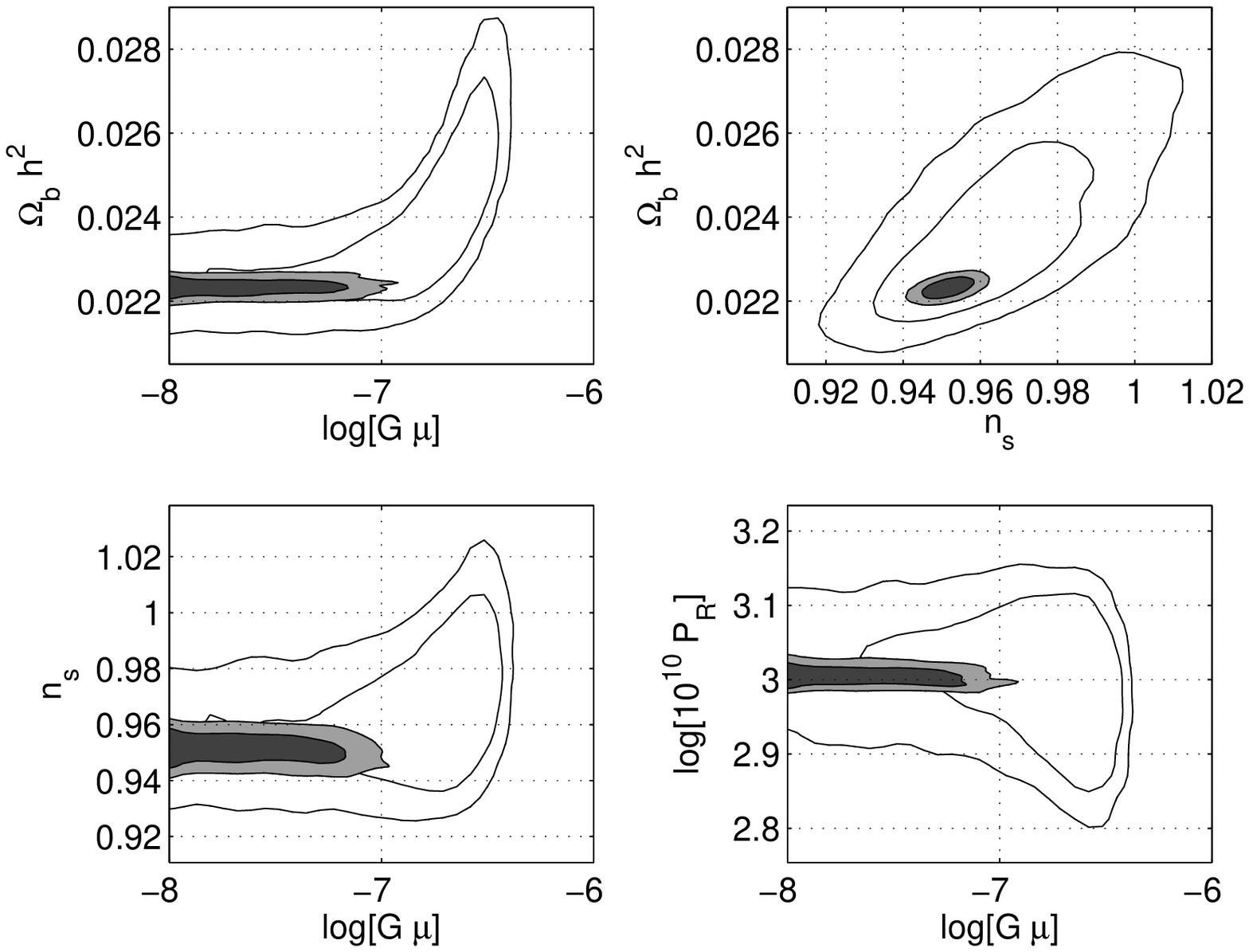,height=4.0in,width=4.0in}
\end{center}
\caption{Results for standard 6 parameter fit with the addition of cosmic strings. The contours show $68\%$ and $95\%$ confidence intervals for WMAP (light contours)  and PLANCK (dark contours). Notice that the degeneracies between $n_{\rm s}$, $\Omega_{\rm b}h^2$ and $G\mu$ which is very obvious for WMAP is broken by the high resolution PLANCK data, allowing each of the parameters to be measured individually.}
\label{figure:planck7}
\end{figure}

The results of this are shown in Fig~\ref{figure:planck7}. For comparison, we have also included equivalent results for WMAP which were obtained in earlier work~\cite{Battye:2006pk}. There is a notable reduction in the range $n_{\rm s}$ can take due to the increased sensitivity and resolution of PLANCK. Furthermore, the degeneracy between $G\mu$ and other parameters, such as $\Omega_{\rm b}h^2$ and $n_{\rm s}$ is broken. For WMAP, we found a $95\%$ confidence upper bound of $ G\mu < 3.0 \times 10^{-7}$; for the simulated PLANCK data this is reduced to  $ G\mu < 6.5 \times 10^{-8}$.

\begin{figure}[t]
\begin{center}
\epsfig{file=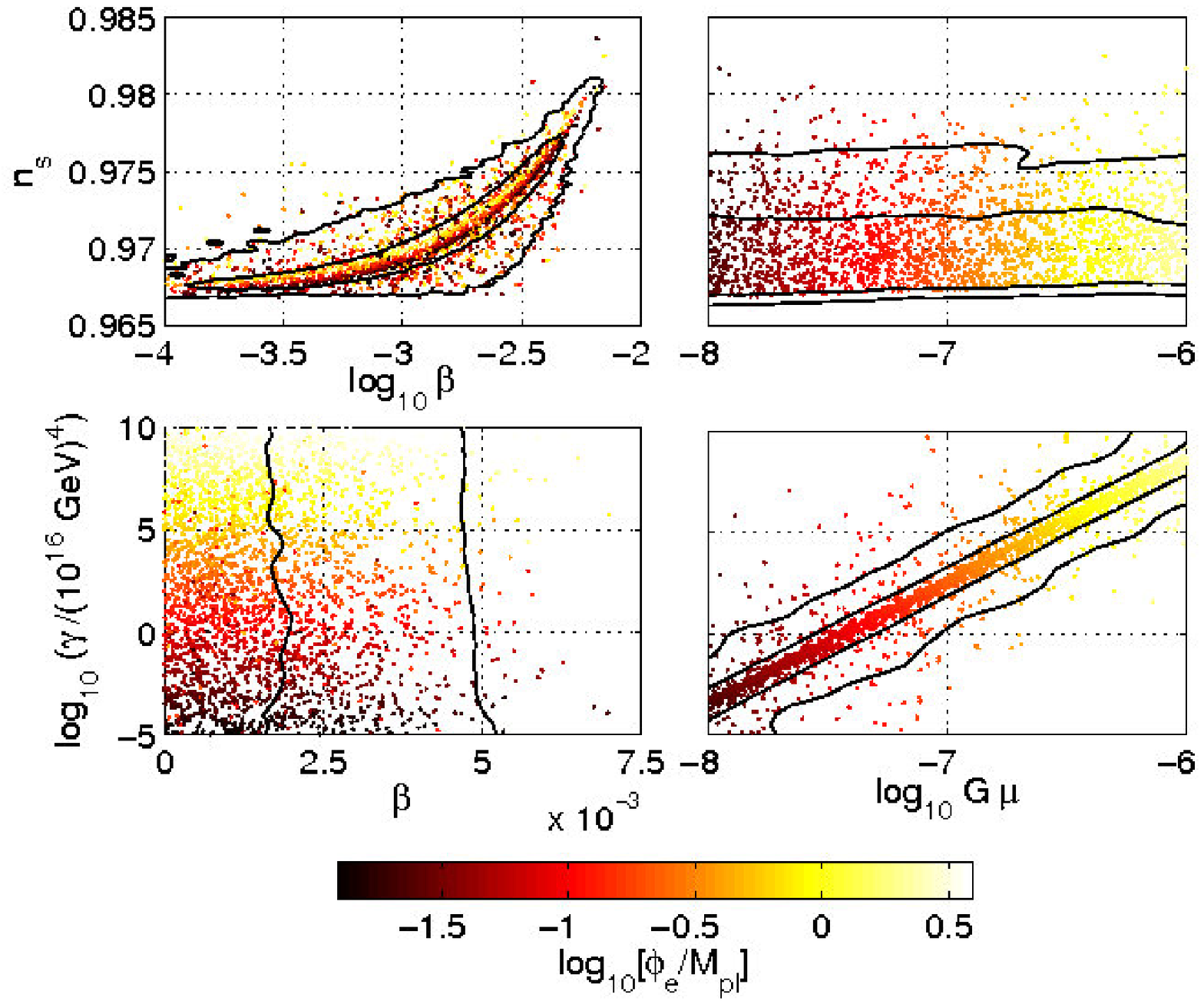,height=3.5in,width=3.5in}
\epsfig{file=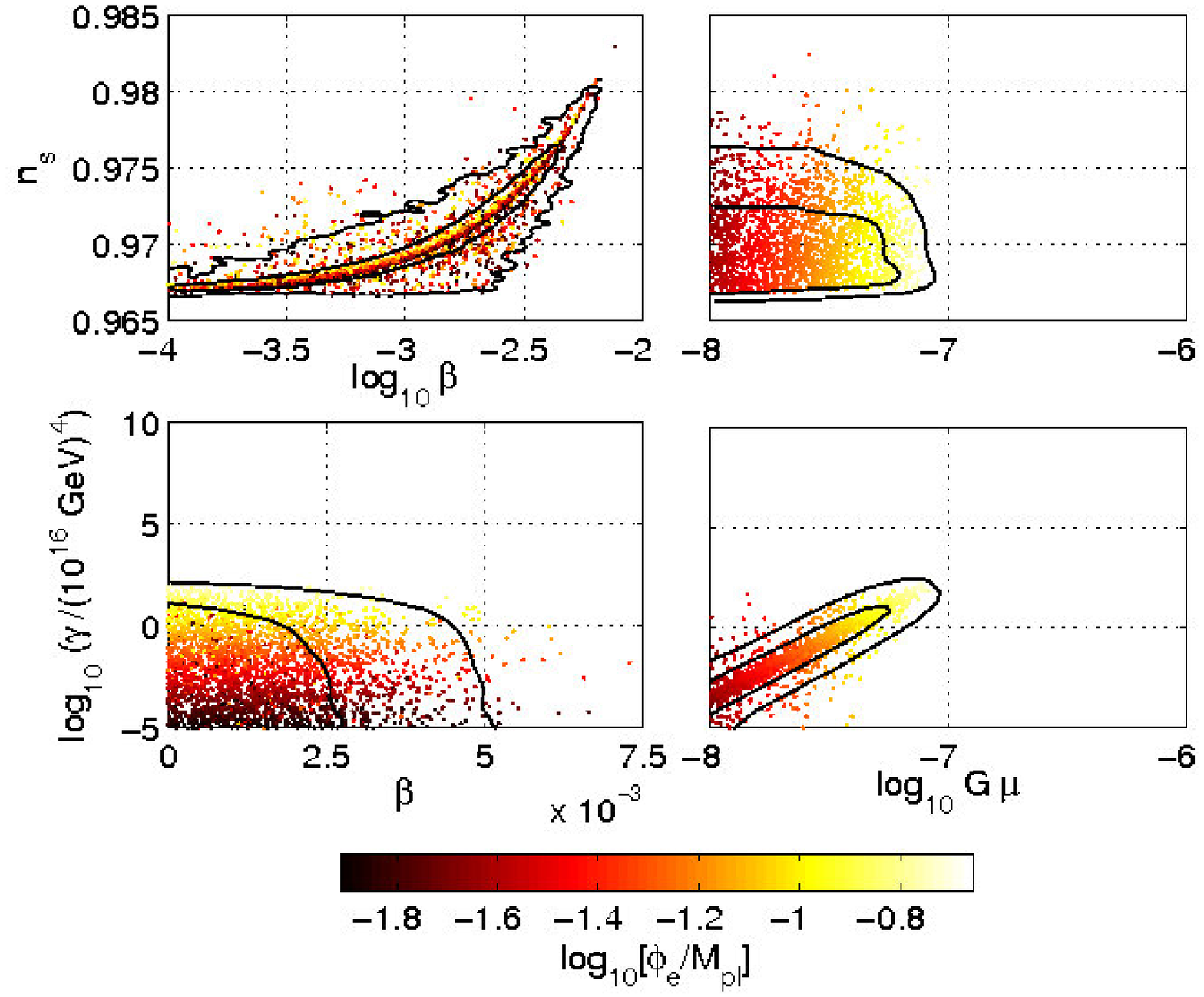,height=3.5in,width=3.5in}
\end{center}
\caption{Same as Fig.~\ref{figure:bi_current} for simulated PLANCK data. Note that the scales are the same as for the current constraints, illustrating substantial improvement likely from PLANCK.}
\label{figure:bi_planck}
\end{figure}

We now assess the improved constraints on brane inflation models using simulated PLANCK data. For the fiducial set of input $C_{\ell}$'s  we use the best fit  parameters from a brane inflation model with $\beta=0$. These parameters are similar to the standard $\Lambda$CDM values used above, with the exception that $n_{\rm s}$ is slightly higher (=0.967). The results of the analysis are shown in Fig.~\ref{figure:bi_planck}. In the case where we do not include the string contribution, the derived value of $\phi_{\rm e}$ can again exceed $M_{\rm P}$. The constraint on $\beta$ is improved due to the tightening of that on $n_{\rm s}$ with the $95\%$ confidence upper limit being reduced to $\beta < 0.004$. Including the string contribution does not modify constraints on $n_{\rm s}$ and $\beta$, but provides more stringent limits on $\gamma$ and $\phi_{\rm e}$. The $95\%$ confidence upper limits for these parameters are reduced to $\log_{10}(\gamma/(10^{16}{\rm GeV})^4)<1.1$ and $\phi_{\rm e}<0.11 \,M_{\rm P}$. The 4 order of magnitude improvement in the constraint on $\gamma$ is due to the fact that $P_{\cal R}\propto \gamma^{-1/6}$.

We note that in addition to constraints from the TT, EE and TE, it is possible to constrain or even detect the effect of cosmic strings using the BB spectrum. It was suggested in ref.~\cite{Battye:2006pk} that maximum amplitude of the power spectrum for a model compatiable with the present data is $\approx 0.3\mu{\rm K}$ at $\ell\sim 1000$. Similar conclusions were found in ref.~\cite{Bevis:2007qz}.

\section{Implications for the scalar-to-tensor ratio} \label{sec:strat}

The most significant implications of these results are for the possible values of $r$ allowed within such scenarios. From the expression for the scalar amplitude (\ref{amplitude}) and the tensor-to-scalar ratio (\ref{scalartensor})
\begin{equation}
{V_0\over M_{\rm P}^4}={3\pi^2\over 2} rP_{\cal R}\,.
\end{equation}
Since $G\mu\propto V_0^{1/2}$ (\ref{stringtension}), 
we see that an upper bound on the string tension is also an upper bound on $V_0$ and therefore on $r$. For $\sqrt{P_{\cal R}}=4.5\times 10^{-5}$, one finds
\begin{equation}
G\mu=4.3\times 10^{-5}\left({r/M}\right)^{1/2}\,,
\end{equation}
where we have reinstated $M$, the number of antibranes, as a parameter. Hence, if there is a limit of order $G\mu<2-3\times 10^{-7}$ imposed by the CMB data, then $r/M< 2\times 10^{-5}$. If we assume, as we have done in the rest of the paper that $M=1$, then there is a limit on the value of $r$ for any brane inflation model of this kind which is well below that which is likely to be observed by even the most ambitious CMB polarization instruments being considered, and it would require large values of $M\gg 100$ to make this possible. Although the argument above applies strictly only to the $\beta=0$, similar arguments can be made for low $\beta$ where inflation takes place in the slow-roll regime.
This is interesting since other model dependent limits on $r$  have been considered in the literature~\cite{Baumann:2006cd,Kallosh:2007wm}. 

For models of brane inflation involving mobile brane moving down 
warped throats, there is another upper limit on $r$ 
which can be derived from the upper bound on the inflaton field. 
The inflaton field is identified by the brane-antibrane separation
and it cannot exceed the geometric limit imposed by size of the 
compactification manifold. One finds that~\cite{Baumann:2006cd}
$(\phi_{\rm e}/M_{P})^2 < 4/n$
where the integer $n$ is the total $D3$ brane charge which generates the
warped geometry. Using the above geometric bound on the values of the inflaton, one can show that $r < 3.6\times 10^{-2}/n$  for warped brane inflation models. Since the supergravity approximation is valid in the large $n$ limit, $n \gg 1$, this seems to predict a very small amplitude for tensor modes. When compared with our bound from cosmic strings $\phi_{\rm e}<0.66 M_{\rm P}$, which applies to the simple model (\ref{simppot}), the above limit is stricter if $n>9$. We conclude, therefore, that our observational limit its compatible with these theoretical arguments.

In the simplest brane inflation model featuring moduli stabilization~\cite{Kachru:2003aw} the value of $r$ can be related to the gravitino mass, $m_{3/2}$. Such a model is essentially a single-field slow-roll model in which inflation occurs along a trough in the potential. The non-perturbative effects 
responsible for stabilizing the volume modulus generate an Anti-de-Sitter minimum, and the de-Sitter minimum is obtained by adding antibranes whose (warped) tension contribute a 
positive term to the overall value of the potential. The presence of the 
antibranes breaks supersymmetry, the breaking scale being set by the brane 
tension. Hence, one finds that the Hubble parameter during inflation is $H^2\propto V/M_{\rm P}^2\propto m_{3/2}^2$, which implies that $V\propto M_{\rm P}^2m_{3/2}^2$, that is the energy scale of inflation, $V^{1/4}$ is the geometric mean of $M_{\rm P}$ and $m_{3/2}$. For the most commonly used gravitino mass $m_{3/2} \sim 1 {\rm TeV}$ the tensor modes are not observable, since this implies $r \sim 10^{-24}$. For these small values of $r$, $V_0^{1/4}<3\times 10^{10} {\rm GeV}$, which is a much stronger bound than the one derived from cosmic strings here. Conversely, if this gravitino bound applies, the observation of tensor modes and cosmic strings will not be feasible in any realistic experiments. We emphasize, however, that not all models are constrained this way. For example in models where the uplifting is achieved by a racetrack superpotential~\cite{Kallosh:2004yh}, the gravitino mass is not related to the uplifting energy, and therefore
the bound discussed above no longer applies. 

\section{Discussion and conclusions} \label{sec:discussion}

In this paper we have presented limits and constraints on the parameters (and derived parameters) of brane inflation models in the slow-roll regime when a cosmic string component is included in the fitting. Important aspects of the results are an increased range of acceptable values of $n_{\rm s}$, limits on $\gamma$, which is related to the volume of the internal space, $\beta$ the inflaton mass parameter and a constraint on the value of $\phi_{\rm e}<M_{\rm P}$ irrespective of any theoretical considerations with the consequent implications for $r$.

We note that one way out of the bounds which we have discussed here is to
consider the large $\beta$ regime, where there are again viable inflationary
models~\cite{Bean:2007hc}. These models fall in the category of inflationary
models with general speed of sound studied in~\cite{Peiris:2007gz}.  In this case fast-roll inflation is possible due to the kinetic term actually being of the Dirac-Born-Infeld type (DBI)~\cite{Silverstein:2003hf,Alishahiha:2004eh,Kecskemeti:2006cg}. This modifies some of the preceding discussion. The Lagrangian for DBI inflation is\be
{\mathcal L} = \frac{1}{f\left(\phi\right)}\left[1-\sqrt{1-
  f\left(\phi\right)g^{\mu\nu}\partial_{\mu}\phi\partial_{\nu}\phi}\right] - 
V\left(\phi\right)\,,
\ee
where the function $f\left(\phi\right)$ depends on the throat geometry and the potential $V(\phi)$ would be given by (\ref{masspot}).

In this case, the expression for the scalar-to-tensor ratio is modified to $r=16c_{\rm s}\epsilon$ where the parameter $\epsilon$ is defined as a generalized slow-roll parameter $\epsilon = 2c_{\rm s}M_{\rm P}^2(H^{\prime}/H)^2$ and the speed of sound is
\be
c_{\rm s}^{2} = 1-
f\left(\phi\right)g^{\mu\nu}\partial_{\mu}\phi\partial_{\nu}\phi\,.
\ee
Since $c_{\rm s}^2\le 1$ one might think that $r$ would be small, however $\epsilon$ can be much larger in fast roll models. This was emphasized in ref.~\cite{Bean:2007hc} who also pointed out in the case where $r$ is small the non-Gaussianity of the density fluctuations, quantified by $f_{\rm NL}$, would be particularly high due to consistency relation~\cite{Lidsey:2006ia}
\be
1-n_{\rm s}=0.4r\sqrt{f_{\rm NL}}\,,
\ee
which was derived in the equilateral triangle limit in momentum space. 

As $\beta$ increases in the slow-roll $\text{regime}^{\footnotemark[1]}$
\footnotetext[1]{Simultaneously with our 
  work, Ref.\cite{Bean:2007eh} appeared where an alternative model,
  the so-called infrared DBI brane inflation, is compared with observations. For this
  class of models an even stronger bound on the cosmic string tension is found,
  $G\mu < 10^{-14}$.} 
the value of $n_{\rm s}$ increases beyond that which is compatible with the data. However, at some value of $\beta$ the effects of the DBI kinetic term kick in and the potential (\ref{masspot}) becomes dominated by the term $\propto\beta$. Hence $P_{\cal R}$ and $n_{\rm s}$ are those of simple Klein-Gordon field, albeit modified by $c_{\rm s}$. This breaks the link between $G\mu$ and $r$, and therefore the constraints discussed in the previous sections only apply in the slow roll regime. 

We note that all the brane inflation models constructed so far require some amount of  fine-tuning to work. In the present case, the tuning corresponds to the value of $\beta$ being low. The effects of moduli stabilization have 
led to a large value of the $\eta$ parameter, making fine-tuning necessary
in models of F-term inflation~\cite{Kachru:2003sx,McAllister:2005mq}.
D-term inflation models do not suffer form the same $\eta$-problem
but threshold corrections to the superpotential \cite{Berg:2004ek,Berg:2005ja}
generate a large mass for the inflaton, making fine-tuning necessary in
these models as well. A possible way to alleviate the fine-tuning problem 
is made possible by a remarkable property of a class of multi-brane
models~\cite{Cline:2005ty}. Usually one balances the effect of the 
volume stabilization mechanism against the 
Coulombic brane-antibrane interaction to obtain a potential with a 
flat enough region to support inflation~\cite{Burgess:2004kv}. The model
requires a fine balancing of the two effects. For generic values of the 
parameters the potential will either be too steep, or it will feature a 
local de-Sitter minimum where the mobile branes being separated from the
antibranes by a potential barrier. However, if one or more branes tunnel
out of the local minimum and annihilates with the antibranes, the height 
of the barrier decreases, and for a critical number of branes it disappears, 
resulting in a monotonic but almost flat potential. Inflation then proceeds
via slow-roll as the remaining branes roll towards the antibranes stuck at
the bottom of the warped throat.  We will investigate this type of model using the techniques applied here in a future study.

A similar, but complementary study, to ours for the case $\beta=0$ but not including the string component has been performed in ref.~\cite{Lorenz:2007ze}. In order to get meaningful constraints they applied the bound on the exit scale as suggested in ref.\cite{Baumann:2006cd} which is discussed in section~\ref{sec:strat}. This study is relevant to the domain where the string tension is weakened by a large number of antibranes. We estimate that if $M>36$ these constraints will apply.

Finally we comment that there may be limits on $G\mu$ which come from pulsar
timing if the cosmic string network achieves scaling by the creation of loops and the subsequent emission of radiation (see ref.~\cite{Caldwell:1996en} and references therein). We caution that these should  be considered to be less robust since they are more strongly effected by the small-scale dynamics, such as loop formation, of the cosmic string network. This is not completely understood in the case of standard cosmic strings in 3+1 dimensions; the situation with respect to the higher dimension cosmic strings is even less clear. Nonetheless, limits on the energy density in gravitational waves, $\Omega_{\rm g}h^2$, have substantially improved in recent times. Probably the most reliable limit is $\Omega_{\rm g}h^2<2\times 10^{-8}$ at frequencies $f=2\times 10^{-9}{\rm Hz}$~\cite{Jenet:2006sv}. Limits on $G\mu$ from such a bound were considered in ref.~\cite{Battye:2006pk} and they are dependent on the loop production size relative to the horizon $\alpha$. If $\alpha<10^{-4}$ then $G\mu>10^{-6}$ is excluded and hence the limit from CMB anisotropy is strongest, whereas for $\alpha>10^{-4}$ one finds that $G\mu> 10^{-10}/\alpha$ is excluded, which would be tighter than the CMB limit for $\alpha>10^{-3}$. It is clear that an improved understanding of the loop production mechanism coupled with expected improvements in the bound on $\Omega_{\rm g}h^2$ could lead to a more powerful constraint than is expected from PLANCK.


\begin{thebibliography}{999}

\bibitem{Smoot:1992td}
  G.~F.~Smoot {\it et al.},
  ``Structure in the COBE differential microwave radiometer first year maps,''
  Astrophys.\ J.\  {\bf 396} (1992) L1.

\bibitem{Spergel:2006hy}
  D.~N.~Spergel {\it et al.}  [WMAP Collaboration],
  ``Wilkinson Microwave Anisotropy Probe (WMAP) three year results:
  Implications for cosmology,''
  Astrophys.\ J.\ Suppl.\  {\bf 170} (2007) 377
  [arXiv:astro-ph/0603449].

\bibitem{Dvali:1998pa}
  G.~R.~Dvali and S.~H.~H.~Tye,
  ``Brane inflation,''
  Phys.\ Lett.\  B {\bf 450} (1999) 72
  [arXiv:hep-ph/9812483].

\bibitem{Burgess:2001fx}
  C.~P.~Burgess, M.~Majumdar, D.~Nolte, F.~Quevedo, G.~Rajesh and R.~J.~Zhang,
  ``The inflationary brane-antibrane universe,''
  JHEP {\bf 0107}, 047 (2001)
  [arXiv:hep-th/0105204].

\bibitem{GarciaBellido:2001ky}
  J.~Garcia-Bellido, R.~Rabadan and F.~Zamora,
  ``Inflationary scenarios from branes at angles,''
  JHEP {\bf 0201}, 036 (2002)
  [arXiv:hep-th/0112147].

\bibitem{C_Strings1}
  A.~Sen,
  ``Tachyon condensation on the brane antibrane system,''
  JHEP {\bf 9808}, 012 (1998)
  [arXiv:hep-th/9805170].

  E.~Witten,
  ``D-branes and K-theory,''
  JHEP {\bf 9812}, 019 (1998)
  [arXiv:hep-th/9810188].

  P.~Horava,
  ``Type IIA D-branes, K-theory, and matrix theory,''
  Adv.\ Theor.\ Math.\ Phys.\  {\bf 2}, 1373 (1999)
  [arXiv:hep-th/9812135].

  M.~Majumdar and A.~Christine-Davis,
  ``Cosmological creation of D-branes and anti-D-branes,''
  JHEP {\bf 0203}, 056 (2002)
  [arXiv:hep-th/0202148].

\bibitem{C_Strings2}
  S.~Sarangi and S.~H.~H.~Tye,
  ``Cosmic string production towards the end of brane inflation,''
  Phys.\ Lett.\  B {\bf 536}, 185 (2002)
  [arXiv:hep-th/0204074].

  N.~T.~Jones, H.~Stoica and S.~H.~H.~Tye,
  ``The production, spectrum and evolution of cosmic strings in brane
  inflation,''
  Phys.\ Lett.\  B {\bf 563}, 6 (2003)
  [arXiv:hep-th/0303269].

  N.~Barnaby, A.~Berndsen, J.~M.~Cline and H.~Stoica,
  ``Overproduction of cosmic superstrings,''
  JHEP {\bf 0506}, 075 (2005)
  [arXiv:hep-th/0412095].
\bibitem{Kachru:2003sx}
  S.~Kachru, R.~Kallosh, A.~Linde, J.~M.~Maldacena, L.~McAllister and S.~P.~Trivedi,
  ``Towards inflation in string theory,''
  JCAP {\bf 0310} (2003) 013
  [arXiv:hep-th/0308055].

\bibitem{Tachyon_Strings}
  P.~Kraus and F.~Larsen,
  ``Boundary string field theory of the DD-bar system,''
  Phys.\ Rev.\  D {\bf 63}, 106004 (2001)
  [arXiv:hep-th/0012198].

  T.~Takayanagi, S.~Terashima and T.~Uesugi,
  ``Brane-antibrane action from boundary string field theory,''
  JHEP {\bf 0103}, 019 (2001)
  [arXiv:hep-th/0012210].




\bibitem{CHMb}
  C.~Contaldi, M.~Hindmarsh and J.~Magueijo,
  ``The power spectra of CMB and density fluctuations seeded by local  cosmic
  strings,''
  Phys.\ Rev.\ Lett.\  {\bf 82} (1999) 679
  [arXiv:astro-ph/9808201].

\bibitem{WB}
  R.~A.~Battye and J.~Weller,
  ``Cosmic structure formation in hybrid inflation models,''
  Phys.\ Rev.\ D {\bf 61} (2000) 043501
  [arXiv:astro-ph/9810203].

\bibitem{Wyman:2005tu}
  M.~Wyman, L.~Pogosian and I.~Wasserman,
  ``Bounds on cosmic strings from WMAP and SDSS,''
  Phys.\ Rev.\ D {\bf 72} (2005) 023513
  [Erratum-ibid.\ D {\bf 73} (2006) 089905]
  [arXiv:astro-ph/0503364].

\bibitem{Seljak:2006bg}
  U.~Seljak, A.~Slosar and P.~McDonald,
  ``Cosmological parameters from combining the Lyman-alpha forest with CMB,
  galaxy clustering and SN constraints,''
  arXiv:astro-ph/0604335.

\bibitem{Bevis:2006mj}
  N.~Bevis, M.~Hindmarsh, M.~Kunz and J.~Urrestilla,
  ``CMB power spectrum contribution from cosmic strings using field-evolution
  simulations of the Abelian Higgs model,''
  arXiv:astro-ph/0605018.

\bibitem{Battye:2006pk}
  R.~A.~Battye, B.~Garbrecht and A.~Moss,
  ``Constraints on supersymmetric models of hybrid inflation,''
  JCAP {\bf 0609} (2006) 007
  [arXiv:astro-ph/0607339].

\bibitem{Bevis:2007gh}
  N.~Bevis, M.~Hindmarsh, M.~Kunz and J.~Urrestilla,
  ``Fitting CMB data with cosmic strings and inflation,''
  arXiv:astro-ph/0702223.

\bibitem{Klebanov:2000hb}
  I.~R.~Klebanov and M.~J.~Strassler,
  ``Supergravity and a confining gauge theory: Duality cascades and
  chiSB-resolution of naked singularities,''
  JHEP {\bf 0008}, 052 (2000)
  [arXiv:hep-th/0007191].

\bibitem{Firouzjahi:2005dh}
  H.~Firouzjahi and S.~H.~Tye,
  ``Brane inflation and cosmic string tension in superstring theory,''
  JCAP {\bf 0503} (2005) 009
  [arXiv:hep-th/0501099].

\bibitem{VP}
  L.~Pogosian and T.~Vachaspati,
  ``Cosmic microwave background anisotropy from wiggly strings,''
  Phys.\ Rev.\ D {\bf 60} (1999) 083504
  [arXiv:astro-ph/9903361].

\bibitem{ABR}
  A.~Albrecht, R.~A.~Battye and J.~Robinson,
  ``The case against scaling defect models of cosmic structure formation,''
  Phys.\ Rev.\ Lett.\  {\bf 79} (1997) 4736
  [arXiv:astro-ph/9707129];
  R.~A.~Battye, J.~Robinson and A.~Albrecht,
  ``Structure formation by cosmic strings with a cosmological constant,''
  Phys.\ Rev.\ Lett.\  {\bf 80} (1998) 4847
  [arXiv:astro-ph/9711336];
  A.~Albrecht, R.~A.~Battye and J.~Robinson,
  ``A detailed study of defect models for cosmic structure formation,''
  Phys.\ Rev.\ D {\bf 59} (1999) 023508
  [arXiv:astro-ph/9711121].

\bibitem{Seljak}
 U.~Seljak and M.~Zaldarriaga,
     ``A Line of Sight Approach to Cosmic Microwave Background Anisotropies,''
     Ap.~J {\bf 469} (1996) 437
     [arXiv:astro-ph/9603033].

\bibitem{LC}
  A.~Lewis, A.~Challinor and A.~Lasenby,
    ``Efficient Computation of {CMB} anisotropies in closed {FRW} models,''
     Ap.~J {\bf 538} (2000) 473
     [arXiv:astro-ph/9911177].

\bibitem{Kachru:2003aw}
  S.~Kachru, R.~Kallosh, A.~Linde and S.~P.~Trivedi,
  ``De Sitter vacua in string theory,''
  Phys.\ Rev.\  D {\bf 68} (2003) 046005
  [arXiv:hep-th/0301240].

\bibitem{McAllister:2005mq}
  L.~McAllister,
  ``An inflaton mass problem in string inflation from threshold corrections  to
  volume stabilization,''
  JCAP {\bf 0602}, 010 (2006)
  [arXiv:hep-th/0502001].

\bibitem{DeWolfe:2002nn}
  O.~DeWolfe and S.~B.~Giddings,
  ``Scales and hierarchies in warped compactifications and brane worlds,''
  Phys.\ Rev.\  D {\bf 67}, 066008 (2003)
  [arXiv:hep-th/0208123].

\bibitem{Burgess:2004kv}
  C.~P.~Burgess, J.~M.~Cline, H.~Stoica and F.~Quevedo,
  ``Inflation in realistic D-brane models,''
  JHEP {\bf 0409}, 033 (2004)
  [arXiv:hep-th/0403119].

\bibitem{Berg:2004ek}
  M.~Berg, M.~Haack and B.~Kors,
  ``Loop corrections to volume moduli and inflation in string theory,''
  Phys.\ Rev.\  D {\bf 71}, 026005 (2005)
  [arXiv:hep-th/0404087].

\bibitem{Berg:2005ja}
  M.~Berg, M.~Haack and B.~Kors,
  ``String loop corrections to Kaehler potentials in orientifolds,''
  JHEP {\bf 0511}, 030 (2005)
  [arXiv:hep-th/0508043].

\bibitem{Berg:2004sj}
  M.~Berg, M.~Haack and B.~Kors,
  ``On the moduli dependence of nonperturbative superpotentials in brane
  arXiv:hep-th/0409282.

\bibitem{Baumann:2007np}
  D.~Baumann, A.~Dymarsky, I.~R.~Klebanov, L.~McAllister and P.~J.~Steinhardt,
  Phys.\ Rev.\ Lett.\  {\bf 99}, 141601 (2007)
  [arXiv:0705.3837 [hep-th]].

\bibitem{Baumann:2007ah}
  D.~Baumann, A.~Dymarsky, I.~R.~Klebanov and L.~McAllister,
  arXiv:0706.0360 [hep-th].

\bibitem{LB}
  A.~Lewis and S.~Bridle,
     ``Cosmological parameters from CMB and other data: a Monte-Carlo approach,''
     Phys.\ Rev.\ D {\bf 66} (2002) 103511
     [arXiv:astro-ph/0205436].

\bibitem{Hinshaw}
     G.~Hinshaw {\it et al.},
     ``Three-Year Wilkinson Microwave Anisotropy Probe (WMAP) Observations: Temperature Analysis,''
     arXix:astro-ph/0603451.

\bibitem{Page}
  L.~Page {\it et al.},
  ``Three Year Wilkinson Microwave Anisotropy Probe (WMAP) Observations: Polarization Analysis,''
     arXix:astro-ph/0603450.

\bibitem{Lyth:1996im}
  D.~H.~Lyth,
  ``What would we learn by detecting a gravitational wave signal in the  cosmic
  microwave background anisotropy?,''
  Phys.\ Rev.\ Lett.\  {\bf 78} (1997) 1861
  [arXiv:hep-ph/9606387].

\bibitem{Efstathiou:2005tq}
  G.~Efstathiou and K.~J.~Mack,
  ``The Lyth Bound Revisited,''
  JCAP {\bf 0505}, 008 (2005)
  [arXiv:astro-ph/0503360].

\bibitem{bluebook}
Planck: The Scientific Programme, ESA-SCI (2005) 1.

\bibitem{Lewis:2006ym}
    A.~L.~Lewis, J.~Weller and R.~A.~Battye,
    	``The Cosmic Microwave Background and the Ionization History
                 of the Universe,''
	MNRAS {\bf 373} (2006) 561 [arXiv:astro-ph/0606552].

\bibitem{Bevis:2007qz}
  N.~Bevis, M.~Hindmarsh, M.~Kunz and J.~Urrestilla,
  Phys.\ Rev.\  D {\bf 76}, 043005 (2007)
  [arXiv:0704.3800 [astro-ph]].

\bibitem{Baumann:2006cd}
  D.~Baumann and L.~McAllister,
  ``A microscopic limit on gravitational waves from D-brane inflation,''
  Phys.\ Rev.\  D {\bf 75} (2007) 123508
  [arXiv:hep-th/0610285].

\bibitem{Kallosh:2007wm}
  R.~Kallosh and A.~Linde,
  ``Testing String Theory with CMB,''
  JCAP {\bf 0704}, 017 (2007)
  [arXiv:0704.0647 [hep-th]].

\bibitem{Kallosh:2004yh}
  R.~Kallosh and A.~Linde,
  ``Landscape, the scale of SUSY breaking, and inflation,''
  JHEP {\bf 0412}, 004 (2004)
  [arXiv:hep-th/0411011].

\bibitem{Bean:2007hc}
  R.~Bean, S.~E.~Shandera, S.~H.~Henry Tye and J.~Xu,
  ``Comparing Brane Inflation to WMAP,''
  JCAP {\bf 0705} (2007) 004
  [arXiv:hep-th/0702107].

\bibitem{Peiris:2007gz}
  H.~V.~Peiris, D.~Baumann, B.~Friedman and A.~Cooray,
  ``Phenomenology of D-Brane Inflation with General Speed of Sound,''
  arXiv:0706.1240 [astro-ph].

\bibitem{Silverstein:2003hf}
  E.~Silverstein and D.~Tong,
  ``Scalar speed limits and cosmology: Acceleration from D-cceleration,''
  Phys.\ Rev.\  D {\bf 70}, 103505 (2004)
  [arXiv:hep-th/0310221].

\bibitem{Alishahiha:2004eh}
  M.~Alishahiha, E.~Silverstein and D.~Tong,
  ``DBI in the sky,''
  Phys.\ Rev.\  D {\bf 70}, 123505 (2004)
  [arXiv:hep-th/0404084].

\bibitem{Kecskemeti:2006cg}
  S.~Kecskemeti, J.~Maiden, G.~Shiu and B.~Underwood,
  ``DBI inflation in the tip region of a warped throat,''
  JHEP {\bf 0609}, 076 (2006)
  [arXiv:hep-th/0605189].

\bibitem{Lidsey:2006ia}
  J.~E.~Lidsey and D.~Seery,
  ``Primordial non-Gaussianity and gravitational waves: Observational tests  of
  brane inflation in string theory,''
  Phys.\ Rev.\  D {\bf 75}, 043505 (2007)
  [arXiv:astro-ph/0610398].

\bibitem{Bean:2007eh}
  R.~Bean, X.~Chen, H.~V.~Peiris and J.~Xu,
  ``Comparing Infrared Dirac-Born-Infeld Brane Inflation to Observations,''
  arXiv:0710.1812 [hep-th].

\bibitem{Cline:2005ty}
  J.~M.~Cline and H.~Stoica,
  ``Multibrane inflation and dynamical flattening of the inflaton  potential,''
  Phys.\ Rev.\  D {\bf 72} (2005) 126004
  [arXiv:hep-th/0508029].

\bibitem{Lorenz:2007ze}
  L.~Lorenz, J.~Martin and C.~Ringeval,
  ``Brane inflation and the WMAP data: a Bayesian analysis,''
  arXiv:0709.3758 [hep-th].

\bibitem{Caldwell:1996en}
  R.~R.~Caldwell, R.~A.~Battye and E.~P.~S.~Shellard,
  ``Relic gravitational waves from cosmic strings: Updated constraints and
  opportunities for detection,''
  Phys.\ Rev.\  D {\bf 54}, 7146 (1996)
  [arXiv:astro-ph/9607130].

\bibitem{Jenet:2006sv}
  F.~A.~Jenet {\it et al.},
  ``Upper bounds on the low-frequency stochastic gravitational wave  background
  from pulsar timing observations: Current limits and future  prospects,''
  Astrophys.\ J.\  {\bf 653}, 1571 (2006)
  [arXiv:astro-ph/0609013].

\end{thebibliography}
\end{document}